\documentstyle[aps,prl,multicol,epsf]{revtex}
\begin{document}

\title{Estimating the Strength of an Elastic Network Using Linear Response}

\author{Gemunu H. Gunaratne}

\address{Department of Physics, University of Houston, Houston, TX 77204}
\address{The Institute of Fundamental Studies, Kandy 20000, Sri Lanka}
\maketitle
\nobreak

\begin{abstract}
Disordered networks of fragile elastic elements have been proposed as a model
for inner porous regions of large bones [Gunaratne et.al., cond-mat/0009221, http://xyz.lanl.gov].
In numerical studies, weakening of such networks is seen to be accompanied by reductions in the fraction
of load carrying bonds. This observation is used to show that 
the ratio $\Gamma$ of linear responses of networks to DC and AC driving can be used as a 
surrogate for their strength. The possibility of using $\Gamma$ as a non-invasive diagnostic 
of osteoporotic bone is discussed.
\end{abstract}

\nobreak
\begin{multicols}{2}

Osteoporosis is a  major socio-economic problem in an aging 
population~\cite{kibAsmi}. Non-invasive diagnostic tools to determine 
the need for therapeutic intervention are essential for effective management 
of the disease. Bone Mineral Density (BMD), or the effective bone density  
is the principal such investigative tool~\cite{cann}.  Ultrasound transmission through 
bone~\cite{njeAhan} and geometrical characteristics of the inner porous 
region or trabecular architecture (TA)~\cite{potAben} are being 
studied as complementary diagnostics. In this Letter, we use results from a 
model system to suggest an additional diagnostic for osteoporosis.

The TA is the principal load carrier in bone of older adults~\cite{njeAhan}, and
cellular models have been used to study TAs~\cite{vajAkra}.  
In Ref. \cite{gunAraj}, a disordered network of fragile elastic elements~\cite{chuAroo}
was proposed as a system to model mechanical properties of a TA. Preliminary
studies are conducted on two dimensional square networks which
include elastic and bond-bending forces. Motivated by conclusions from mechanical 
studies of bone~\cite{hogAruh}, the springs are assumed to satisfy 
a strain-based fracture criterion; specifically the fracture strains of 
elastic elements are chosen from a Weibull distribution~\cite{leaAdux} (with 
parameters $\gamma_e$ and $m$). Bonds are assumed to fracture when changed beyond 
a critical angle; these fracture-angles are chosen from a second Weibull distribution
(with parameters $\gamma_b$ and $m$). Osteoporosis is modeled by random removal of 
a fraction $\nu$ of springs from the network.

The characteristic introduced below includes response of a network to an AC strain,
which depends on the mass distribution on the network. It is modeled by placing 
masses $m$ at the vertices. Viscous effects 
of the surrounding medium are modeled by a 
dissipative force proportional to the speed of each mass. In studies reported here,
points located on the sides were constrained to move vertically.
Numerical studies of the system support the conjecture that these elastic networks 
are a suitable model to study mechanical properties of bone~\cite{gunAraj}. 

\begin{figure}
\epsfxsize=2.0truein
\hskip 0.60truein
\epsffile{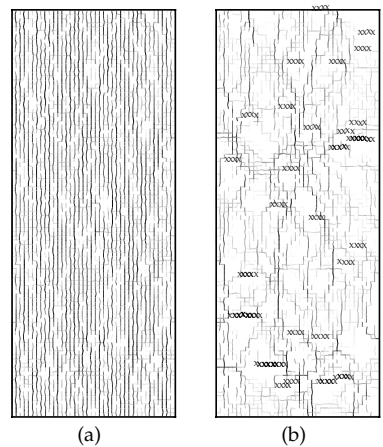}
\vskip 0.05in
\caption{The stress distributions on networks of size $40\times 100$ with  
(a) $\nu=10\%$, and (b) $\nu=30\%$ representing ``healthy" and 
``osteoporotic" bone respectively. For clarity only the compressed bonds 
are shown, and darker hues represent larger stresses. The crosses denote
locations of long horizontal fractures.}
\label{latt}
\end{figure}

Figure \ref{latt} shows the stress distributions 
on two networks subjected to uniform compression~\cite{para,num1}. 
For clarity only the compressed bonds are shown, and darker hues
represent larger stresses. On the ``healthy" network (where $\nu=10\%$)
the large stresses supporting propagation are seen to be distributed
evenly over the network. In contrast, elastic elements supporting a 
``weak" network ($\nu=30\%$)  form a few coherent pathways. 
We refer to the set active in load transmission as the ``stress 
backbone" of a network~\cite{mouAdux}. 

In Figure~\ref{latt}, the X's denote long horizontal fractures; specifically,
locations where four or more consecutive vertical bonds are missing.
It is clear that these fractures prevent the participation of many bonds in 
the stress backbone. Since the number of long fractures increases with $\nu$, 
a progressively smaller fraction of bonds are able to be load carriers. The 
assumption that this dilution of the stress backbone is 
related to the ultimate (or breaking) stress of the network motivates the
measure introduced below.

Consider first an ordered $N\times M$ square network of identical springs, each 
of whose breaking stress, breaking strain and elastic modulus are denoted
by $u_0$, $\zeta_0$ and $Y=u_0/\zeta_0$. Then the ultimate stress and strain of the 
pure (i.e., $\nu=0$) network are $U(0)=N u_0$ and $M \zeta_0$. Denote
by $U(\nu)$ and $\zeta(\nu)$ their values for a network obtained
by removing bonds with a probability $\nu$. Each horizontal layer of the 
network is approximately compressed by $\frac{\zeta(\nu)}{M}$ and a fracture 
will propagate when one of the bonds bordering it is stressed beyond $u_0$. 
In this approximation, the DC response of the network is
\begin{equation}
\chi_0 = \frac{U(\nu)}{\zeta(\nu)} \approx \frac{N Y}{M} \cdot u(\nu),
\label{DCres}
\end{equation}
where $u(\nu) = U(\nu)/U(0)$.

\begin{figure}
\epsfxsize=2.00truein
\hskip 0.60truein
\epsffile{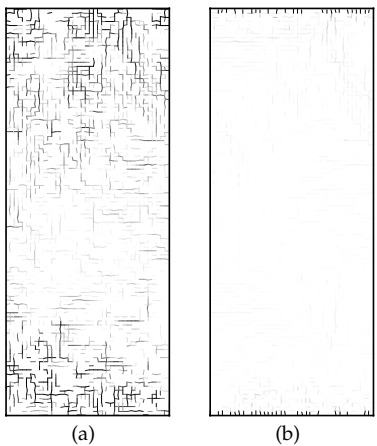}
\caption{The distribution of stresses on the disordered network 
of Figure 1(b) due to small amplitude AC compressions with (a) $\Omega=10$ 
and (b) $\Omega=500$. As $\Omega$ increases, so does the attenuation of the signal, and a progressively 
thinner slice of the network is effected. For sufficiently large $\Omega$ 
only bonds belonging to the top and bottom layers experience an AC stress.}
\label{attn}
\end{figure}

Next we argue that the linear response to an external AC strain 
can be used to estimate the number of bonds in a network.
First subject the network to a DC compression ($\zeta_{DC}$) 
below the yield point, so that there is no fracture of elastic elements. 
Next introduce an additional AC
compression, given by $\zeta(t) = \zeta_{AC} \exp(i\Omega t)$, where 
$\zeta_{AC}\ll \zeta_{DC}$~\cite{num2}. 
When $\Omega$ increases so does the attenuation of the signal, and 
a progressively thinner slice of the network is effected by the AC signal. 
Figure~\ref{attn} shows the AC response of the network of Fig.~\ref{latt}(b)
driven at two frequencies.

\begin{figure}
\epsfxsize=2.50truein
\hskip 0.3truein
\epsffile{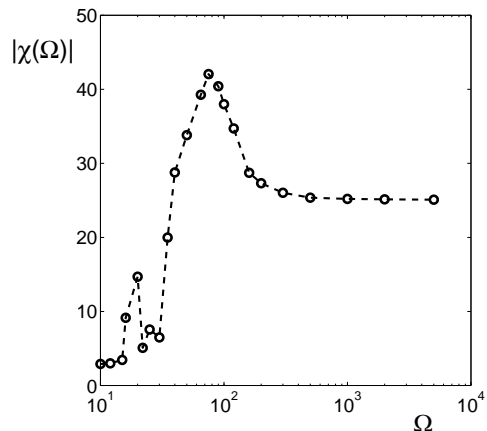}
\caption{The dependence of $|\hat\chi(\Omega)|$ on $\Omega$ for a network with 
$\nu=30\%$. For $\Omega \rightarrow 0$, $\hat\chi(\Omega) \rightarrow \chi_0$,
while for sufficiently large $\Omega$, $\hat\chi(\Omega) \approx N (1-\nu) \bar Y$. 
In this limit, only bonds bordering the 
top and bottom layers experience AC stress. The fluctuations for intermediate
values of $\Omega$ are configuration dependent.}
\label{omgvschi}
\end{figure}

Denote by $T(t)$ the sum of vertical forces on the top layer due to the AC strain.  
The linear response of the network $\hat\chi(\Omega)$ is given by 
$\hat T(\Omega)=\hat\chi(\Omega) \cdot \hat\zeta(\Omega)$, where 
$\hat\zeta(\Omega)$ and $\hat T(\Omega)$ are the Fourier 
transforms of $\zeta(t)$ and $T(t)$ respectively. 
Figure~\ref{omgvschi} shows the behavior of $|\hat\chi(\Omega)|$ for a disordered 
network with $\nu=30\%$. When $\Omega \rightarrow 0$, there is no attenuation and hence
$\hat\chi(\Omega)$ approaches $\chi_0$. On the other hand, 
for sufficiently large $\Omega$ only the two edges are excited and hence
$\hat\chi(\Omega)$ can be expected to approach a limit.
Then each of the $(1-\nu) N$ bonds of the top layer are strained
by a same amount and hence 
\begin {equation}
\hat\chi (\Omega) \approx N (1-\nu) {\bar Y},
\label{ACres}
\end{equation}
where $\bar Y$ denotes the mean elastic constant for bonds on the top layer.
Numerical integrations of disordered networks confirm these conclusions, 
see Fig.~\ref{omgvschi}.

Consider once again the ordered pure network. When only the top and bottom layers 
experience AC stress the effective height of the layer is reduced by a factor $M$; 
thus, $\hat\chi(\Omega)$ will be larger than $\chi_0$ by a factor $M$. For $\nu=0$,
this can be expected to hold (approximately) even for disordered networks since
the stress backbone covers the entire network.  
Even though both $\chi_0$ and $\hat\chi(\Omega)$ will decrease with increasing $\nu$, the
latter will be effected less because long fractures in the middle of the network (which 
reduce the stress backbone) have no effect on $\hat\chi(\Omega)$. 
These arguments motivate the use of  
\begin{equation}
\Gamma(\nu)\equiv \lim_{\Omega \rightarrow \infty} M \frac{\chi_0}{|\hat\chi(\Omega)|}
\label{defn}
\end{equation}
to estimate the reduction in the extent of the stress backbone.
Approximating $\bar Y$ by $Y$ and using Eqs. (\ref{DCres}), (\ref{ACres}) and (\ref{defn}) 
gives
\begin{equation}
u(\nu) \approx (1-\nu)\cdot \Gamma(\nu).
\label{reln3}
\end{equation}

To complete the derivation, we propose a relationship $\nu = F(u)$.
Since $u(\nu)$ for a 2D square network vanishes at the
percolation threshold $\nu=\nu_0=\frac{1}{2}$,
\begin{equation}
\nu_0 = F(0).
\label{limt}
\end{equation}
A form for $F(x)$ has been presented in Ref.~\cite{kahAbat}, where it
was assumed that the propagation of the longest horizontal fracture causes the
collapse of a network. The ingredients used are 
(1) a horizontal fracture of size $k$ enhances the stress on the bordering
elements by $(1 + a k^{\alpha})$, and (2) the size $k_m$ of the longest fracture
satisfies $MN \nu^{k_m} (1-\nu)^2$~\cite{kahAbat}. Since the possible two dimensionality
of a fracture is ignored, the resulting expression does not satisfy Eq.~(\ref{limt}).
We propose a modification
\begin{equation}
\frac{1}{u(\nu)} - 1 \approx a \left( \frac{\ln MN}{\ln (\nu_0/\nu)}\right)^{\alpha},
\label{strength}
\end{equation} 
which agrees with results from numerical integration of disordered networks for
$\nu \in (0, 0.4)$~\cite{oned,perco}. Parameters $a$ and $\alpha$ are expected to
depend on factors such as the relative strengths of the elastic and bond-bending forces.  

Using Eqs.~(\ref{reln3}) and (\ref{limt}), 
\begin{equation}
u(\nu) = (1-\nu_0) \Gamma + h(\Gamma),
\label{main}
\end{equation}
where the nonlinear correction $h(\Gamma)$ can be obtained by inverting (e.g., Born 
expansion) Eq. (\ref{strength}) and will depend on $a$, $\alpha$, and $\ln (MN)$. 

\begin{figure}
\epsfxsize=2.50truein
\hskip 0.30truein
\epsffile{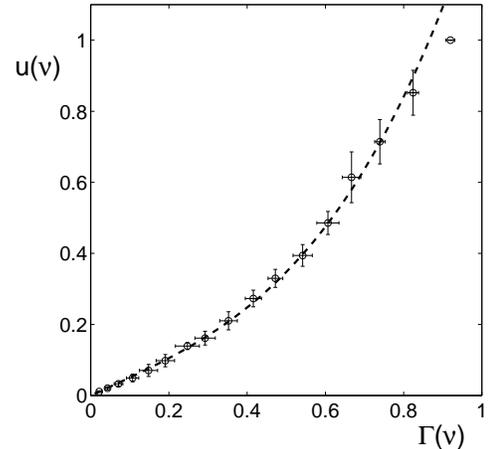}
\caption{The relationship between the function $\Gamma(\nu)$ and $u(\nu)$ 
for a set of 5 equivalent disordered networks. The dashed line shows the
best fit of the data to Eq.~(\ref{main}) with $h(\Gamma)=c\Gamma^{\beta}$.
For these parameters~[10],  $c=0.90$ and $\beta=3.20$.}
\label{gmvsu}
\end{figure}

To test the validity of Eq.~(\ref{main}), numerical studies were conducted on 
a group of five equivalent disordered networks~\cite{para}. 
$\Gamma(\nu)$ and $U(\nu)$ for a given network 
is evaluated using methods discussed earlier~\cite{num1,num2}.
As can be expected, the normalization of $U(\nu)$ by $U(0)$ reduces the variability
between distinct configurations in the group.
In evaluating the linear response, the signal $T(t)$ was collected
after the transients have settled.
For a given $\nu$, there is scatter in the values of both $\Gamma(\nu)$ and
$u(\nu)$, and Figure~\ref{gmvsu} shows their mean and standard errors.
The dashed line is the best fit to the data in the form 
(\ref{main}) with $h(\Gamma) = c\Gamma^{\beta}$. Numerical studies on configurations
with different parameters show that, unlike the linear term of Eq.~(\ref{main}), 
the coefficients $\beta$ and $c$ are parameter dependent.

The reduction of bone strength from its peak value determines the 
level of osteoporosis. Unfortunately, it is not accessible
in-vivo (without breaking a bone!), and surrogates such as 
bone density are used to identify osteoporotic bone. 
The BMD of a patient is compared with that of a sample population
to determine if and when therapeutic interventions are necessary.  However the ultimate 
stress is known to depend on other factors of bone including the structure
of its TA and the ``quality" of bone material. The resulting variations 
make it difficult to identify individuals susceptible
to fracture using measurements of BMD alone~\cite{marAjoh}.

In this Letter, we have shown that linear response of a network 
can be used as a surrogate for its strength. This conjecture is based
on the assumption that the strength is related to the extent of 
the stress backbone. Measurements required to evaluate $\Gamma$ can be
implemented on ex-vivo bone samples. DC strain can be imposed using pressure loading,
and protocols using ultrasonic techniques have been developed to evaluate 
the response of bone samples to AC driving~\cite{couAhob}. Previous studies suggest that 
when their frequencies larger than $\sim 1.5MHz$, ultrasonic signals will excite 
only those trabeculae on the outer edges of a TA~\cite{strAeva}. 
How these measurements can be implemented in-vivo remains to be studied.

Several issues need to be reiterated. To calculate $\hat\chi(\Omega)$,
only elastic forces (on vertices of the top layer) were used. In
driving a bone sample with an AC strain, the matter (on the outer layer)
is accelerated in a dissipative medium. These inertial and 
dissipative forces are proportional to $\Omega^2$ and $\Omega$ respectively.
In contrast, for sufficiently large $\Omega$, $\hat\chi(\Omega)$ is 
$\Omega$-independent (see Fig.~\ref{omgvschi}). Hence, the latter can be 
extracted from the response of the TA to AC signals of several
frequencies.  Secondly, observe that for weak networks ($\nu$ close to $\nu_0$) the
relationship between $u(\nu)$ and $\Gamma(\nu)$ depends only on $\nu_0$.
For smaller values of $\nu$, however, the nonlinear correction $h(\Gamma)$,
which depends on model parameters, becomes relevant. Hence the form of $u(\Gamma)$
may have to be determined for distinct bone locations (e.g., femur, vertebrae)
before a complete diagnostic tool for osteoporosis is developed.
Finally, notice that the definition of $\Gamma$ includes the number 
of layers $M$ of the network. Since the lengths of trabeculae from specific 
anatomical locations are known (typically $\sim$ 1mm), the length of the 
sample can be used to determine $M$.

The author would like to thank S. R. Nagel for pointing out that the 
response of a network is related to its stress backbone.
He also acknowledges discussions with M. P. Marder, G. F. Reiter  
and S. J. Wimalawansa. This research is partially funded by the National Science Foundation, 
the Office of Naval Research and the Texas Higher Education Coordinating
Board.

\end{multicols}

\begin{thebibliography}{99}
\bibitem{para} These networks were obtained by randomly displacing vertices 
  of a square lattice of sides 10 units by up to 1 unit.
  The elastic and bond-bending spring constants were chosen from uniform
  distributions in  [0.5, 1.5] and [2.5, 7.5] respectively.  The fracture criteria 
  are given by $\gamma_e=5\%$, $\gamma_b=0.1$ and $m=5$. Each mass was $10^{-5}$ units,
  and the dissipation coefficient $\eta=10^{-5}$. The disorder of the network is 
  controlled by a single random seed, which is varied to produce additional
  ``equivalent" networks.
\bibitem{num1} The equilibrium was evaluated using the conjugate gradient
  method. The external strain was increased
  in small steps and the network allowed to equilibrate prior to 
  incrementing the strain.
\bibitem{oned} The expression derived in Ref.~\cite{kahAbat} is identical to
Eq.~(\ref{strength}) with $\nu_0$ replaced by the
bond percolation threshold (=1) for a network with a single horizontal layer.
\bibitem{perco} This form is not expected to be valid close to the 
  percolation threshold, see Ref.~\cite{kahAbat}.
\end{thebibliography}
\end{document}